

\def\Version{{4.4}}





 \message{<<Assuming 8.5" x 11" paper>>}

\magnification=\magstep1	          
\raggedbottom
\overfullrule=0pt 

%

\parskip=9pt

\def\singlespace{\baselineskip=12pt}      
\def\sesquispace{\baselineskip=16pt}      










 at10pt
 %

 %
 %
 %










\def\tr{\mathop {{\rm \, Tr}} \nolimits}	 


%

%

%
%



%




\def\sqr#1#2{\vcenter{
  \hrule height.#2pt 
  \hbox{\vrule width.#2pt height#1pt 
        \kern#1pt 
        \vrule width.#2pt}
  \hrule height.#2pt}}




\def\lto{\mathop
        {\hbox{${\lower3.8pt\hbox{$<$}}\atop{\raise0.2pt\hbox{$\sim$}}$}}}
\def\gto{\mathop
        {\hbox{${\lower3.8pt\hbox{$>$}}\atop{\raise0.2pt\hbox{$\sim$}}$}}}
%
%
%



\def\part{\subseteq}		




\def\to{\mathop\rightarrow}	




\def\interior #1 {  \buildrel\circ\over  #1}     




\def\basisvector#1#2#3{
 \lower6pt\hbox{
  ${\buildrel{\displaystyle #1}\over{\scriptscriptstyle(#2)}}$}^#3}



\def\hat{\widehat}		







\fontdimen16\textfont2=2.5pt
\fontdimen17\textfont2=2.5pt
\fontdimen14\textfont2=4.5pt
\fontdimen13\textfont2=4.5pt 

\let\miguu=\footnote
\def\footnote#1#2{{$\,$\parindent=9pt\baselineskip=13pt%
\miguu{#1}{#2\vskip -7truept}}}
 %

\def\linebreak{\hfil\break}
\def\lbr{\linebreak}
\def\pagebreak{\vfil\break}


\def\BulletItem #1 {\item{$\bullet$}{#1 }}
\def\bulletitem #1 {\BulletItem{#1}}

\def\PrintVersionNumber{
 \vskip -1 true in \medskip 
 \rightline{version \Version} 
 \vskip 0.3 true in \bigskip \bigskip}

\def\author#1 {\medskip\centerline{\it #1}\bigskip}

\def\address#1{\centerline{\it #1}\smallskip}

\def\furtheraddress#1{\centerline{\it and}\smallskip\centerline{\it #1}\smallskip}

\def\email#1{\smallskip\centerline{\it address for email: #1}} 

\def\AbstractBegins
{
 \singlespace                                        
 \bigskip\leftskip=1.5truecm\rightskip=1.5truecm     
 \centerline{\bf Abstract}
 \smallskip
 \noindent	
 } 
\def\AbstractEnds
{
 \bigskip\leftskip=0truecm\rightskip=0truecm       
 }

\def\section #1 {\bigskip\noindent{\headingfont #1 }\par\nobreak\smallskip\noindent}

\def\subsection #1 {\medskip\noindent{\subheadfont #1 }\par\nobreak\smallskip\noindent}
 %

\def\ReferencesBegin
{
 \singlespace					   
 \vskip 0.5truein
 \centerline           {\bf References}
 \par\nobreak
 \medskip
 \noindent
 \parindent=2pt
 \parskip=6pt			
 }
 %


\def\reference{\hangindent=1pc\hangafter=1} 

\def\ref{\reference}

\def\sepref{\parskip=4pt \par \hangindent=1pc\hangafter=0}
 %

\def\journaldata#1#2#3#4{{\it #1\/}\phantom{--}{\bf #2$\,$:} $\!$#3 (#4)}
 %

\def\eprint#1{{\tt #1}}

 %


\def\webhome{{\tt http://www.pitp.ca/personal/rsorkin/}}
 %

 %



\def\webtilde{\lower2pt\hbox{${\widetilde{\phantom{m}}}$}}

 %

\def\hpf#1{\webhome{\tt{some.papers/}}}
 %

\def\hpfll#1{\webhome{\tt{lisp.library/}}}
 %



\font\titlefont=cmb10 scaled\magstep2 

\font\headingfont=cmb10 at 12pt
%

\font\subheadfont=cmssi10 scaled\magstep1 
%


\font\csmc=cmcsc10  





\def\thesis#1{\noindent {\csmc Thesis} {\it #1}}


\def\journaldata#1#2#3#4 {#1, {#2}, #3.}
\def\Journaldata#1#2#3#4
    {{\it #1\/}\phantom{--}{\bf #2$\,$:} $\!$#3 (#4), } 


\phantom{}

\PrintVersionNumber

%


\PrintVersionNumber

\sesquispace

\centerline{{\titlefont Ten Theses on Black Hole Entropy}\footnote{$^\star$}%
{ Published in 
 \Journaldata{Studies in History and Philosophy of Modern Physics}{36}{291-301}{2005}
 \eprint{hep-th/0504037} 
 \lbr
 \eprint{http://www.pitp.ca/personal/rsorkin/some.papers/118.ten.theses.pdf}
 }}

\bigskip


\singlespace			        

\author{Rafael D. Sorkin}

\address {Perimeter Institute, Waterloo ON, N2L 2Y5, Canada}

\furtheraddress
 {Department of Physics, Syracuse University, Syracuse, NY 13244-1130, U.S.A.}

\email{rsorkin@perimeterinstitute.ca}

\AbstractBegins
   I present a viewpoint on black hole thermodynamics according to which
   the entropy: derives from horizon ``degrees of freedom''; is finite
   because the deep structure of spacetime is discrete; is ``objective''
   thanks to the distinguished coarse graining provided by the horizon;
   and obeys the second law of thermodynamics precisely because the
   effective dynamics of the exterior region is not unitary. 
\AbstractEnds


\sesquispace


Probably few people doubt that the twin phenomena of black hole entropy and
evaporation hold important clues to the nature of quantum spacetime, but
the agreement pretty much ends there.  Starting from the same evidence, different
workers have drawn very different, and partly contradictory, lessons.
On one hand, there is perhaps broad agreement that the finiteness of the
entropy points to an element of discreteness in the deep structure of
spacetime.  On the other hand there is sharp disagreement over whether
the thermal nature of the Hawking radiation betokens an essential
failure of unitarity in quantum gravity or whether it is instead
betraying the need for a radical revision of the spacetime framework, as
contemplated for instance in the ``holographic principle''.  These
alternatives are not necessarily in contradiction, of course, but in
practice, the wish to retain unitarity has been one of the strongest
motivations for taking seriously the latter type of possibility.  My own
belief is that non-unitarity is probably inevitable in connection with
gravity and that, rather than shunning this prospect, we ought to
welcome it because it offers a straightforward way to understand why the
law of entropy increase continues to hold in the presence of event
horizons.  This conclusion is part of an overall viewpoint on black hole
entropy that I believe to be, if not true, then at least
coherent.  In the following lines, I will try to convey this viewpoint
as succinctly as possible
by expressing it as a series of ``theses''.
In other places, I have written on most of these points in more detail.
Here I try only to bring them together and to indicate briefly the
reasoning behind them.
Following each thesis, a handful of relevant references is indicated
in a footnote. 

\thesis{1. The most natural explanation of the area law is that $S$ resides on the horizon.}

By the area law, I mean of course the fact that the entropy that enters
into the ``generalized second law'' for systems including black holes is
proportional to the total area of all horizons which are present.  (More
correctly, the entropy is, in $3+1$-dimensions, associated with a
spacelike or null 3-surface $\Sigma$, and the area in question is that
of the 2-surface in which $\Sigma$ meets the horizon(s).)  I remember
that when I first learned of this law three decades ago, its most
natural interpretation seemed to me to be that the horizon carried some
kind of information with a density of approximately one bit per unit
area, the area being rendered dimensionless by setting to unity the
rationalized gravitational constant $8{\pi}G$.  (The precise formula for
the entropy in these units is $S=2{\pi}A$, where the coefficient $2\pi$
can be interpreted geometrically as the circumference of the unit circle
in the Wick rotated time direction.)\footnote{$^\dagger$}
{These formulas also take Boltzmann's constant to be unity, of course.}
In a simple-minded model, one might picture the associated horizon
degrees of freedom as plaquettes of horizon-surface on which tiny
`1's and `0's are engraved, leading to a total number of configurations
of the order of $N\sim2^A$, whose logarithm yields an entropy of
$S={\log\,}N{\sim\,}A{\,\log\,}2$.  No one would take such a picture literally,
but it can serve to indicate the kind of thing one is looking for, and
this brings me to my second thesis.\footnote{$^\flat$}
{Some references for the first thesis are
 35, 
 5, 
 41, 
 19, 
 45, 
 51,
 1, 
 10,
 18,
 50.}

\thesis{2. What these bits of information represent depends on the deep
           structure of spacetime, most naturally conceived of as discrete.}

I don't believe that this thesis requires much elaboration, but some
reference to an analogous situation in condensed matter physics might
be suggestive, namely a dilute gas at high temperature.  In that case
one knows that the entropy is in effect just counting the total number
of constituents (molecules) of the gas.  Indeed $S$ is precisely
this number modulo a logarithmic pre-factor that depends on the
temperature, the pressure, the molecular mass, and $\hbar$.
The deep structure in this case is just the atomic structure of matter,
and the finiteness of the number of atoms is then what allows the
entropy itself to be finite.  One might expect something similar for the black
hole, where however the ``atoms'' would be constituents of spacetime
rather than constituents of ponderable matter.  The horizon then could
also have ``constituents'' which could play the role of ``molecules''
of the gas.  (See [12] for how such an explanation might play out in
causal set theory.)

It's worth remarking, however, that although the {\it ultimate} degrees
of freedom responsible for the entropy of a gas are discrete, that does
not necessarily prevent one from giving an approximate continuum account
of its entropy in terms of effective fluid degrees of freedom like the
local mass density.  (I don't know whether anyone has tried to do this,
but if not then I think it would make an interesting problem.  See
[44] for some preliminary results in that direction.)  Similarly, it
might be [35] [5] that fluctuations in horizon shape could be taken to be
the main source of the entropy of a black hole.  (See also the comments
on entanglement entropy below.)\footnote{$^\star$}
{Some references for the second thesis are
 6, 
 41, 
 45, 
 44, 
 12, 
 49, 
 35,
 31,
 22.}

\thesis{3. Two main tasks need to be accomplished then:\lbr
     \phantom{mm}\quad $\bullet$ Identify and count the ``bits'' \lbr
     \phantom{mm}\quad $\bullet$ Explain why S increases}


If we want to explain black hole thermodynamics in terms of some
``informational bits'' residing on the horizon, we have to identify
their material basis and then use this knowledge to figure out how to
count them.  It is equally important, however, that we explain why the
total entropy --- that of the horizons plus that of gravity and matter in
the exterior region --- continues to respect the second law.  Although
this second task is often given short shrift, it is in my opinion the
decisive one; for no entropy merits the name unless it can be shown to
increase with time.\footnote{$^\dagger$}%
{Some references for the third thesis are 
 37, 
 5, 
 41, 
 45.}%

\thesis{4. The idea that the degrees of freedom are inside the black hole is wrong}

In contradiction to thesis 1 above, it is sometimes maintained that the
entropy of a black hole refers to degrees of freedom of the interior
region, rather than its boundary, which by definition is the horizon.
If this were so, however, it is difficult to see how one could ever hope
to account for the generalized second law, since the familiar statistical
mechanical derivations of entropy increase presuppose conditions which
are very different from what one expects to find inside a black hole.

To start with, the form of the ``first law'' for stationary black holes
is that appropriate to a subsystem in internal equilibrium, whose gross
features can be characterized by a small number of thermodynamic
parameters (for a Kerr black hole, just the energy and angular
momentum).  The entropy then counts the number of microstates that
contribute to the given macroscopic equilibrium state.  Now, such a
characterization is completely appropriate to the black hole as viewed
from the outside, but it is obviously a very poor match to the interior
region, which is neither in equilibrium (since it is collapsing rapidly)
nor described by a unique macroscopic state.  
Similarly, derivations of entropy increase usually appeal to some form
of ergodicity (at least to justify counting all of the compatible
microstates in the expression for the entropy), whereas conditions in
the interior are no more ergodic than they are stationary.  Rather, if
the classical regime is any guide, the evolution has much more of a
one-way than an ergodic nature.  
And the familiar derivations also take for granted that
differences in the microstate of the subsystem will eventually make
themselves felt in the environment, so that the {\it combined} system of
subsystem plus environment can explore the entire ``phase space''
available to it.  (For example, the loss of ``phase space volume''
associated with a decrease of the subsystem's energy has to be
compensated by an increase in the number of microstates traversed by the
environment).

In this connection, they also assume that the coupling between subsystem
and environment is weak, so that the entropy will be approximately
additive.  Once again, the contrast with the black hole case could
hardly be greater, since there the $outside \to inside$ coupling is very
strong, while the reciprocal coupling, $inside \to outside$, is
non-existent classically (and perhaps exponentially small or zero in the
quantum theory).
In the absence of such a coupling, it is hard to see how {\it anything}
referring to the interior could play a role in the generalized second
law, since the latter pertains entirely to features visible in the
exterior region.

Path integral computations of the entropy also indicate, in their own
way, that the source of the entropy should not be sought beyond the
horizon.  In fact, the Euclidean-signature black hole metrics that enter
into these calculations do not even possess regions corresponding (under
Wick rotation of the time direction) to the interior of the Lorentzian
black hole.\footnote{$^\flat$}
{Some references for the fourth thesis are
 45,
 5,
 41,
 19,
 18,
 34,
 2,
 52,
 33,
 14,
 54,
 22.}

\thesis{5. The dissipative nature of the horizon precludes an appeal to
           unitarity in proving that the entropy increases.}

If we accept that $S$ is a surface effect (not an attribute of the black
hole interior), then the objections I have just recited lose much of
their force, since the horizon {\it is} in equilibrium (for a stationary
black hole) and its evolution is not so obviously non-ergodic.  Moreover
it is only ``marginally decoupled'' from the exterior classically, and
so might in fact be weakly coupled in a suitable sense in the quantum
theory.  A traditional proof of the second law utilizing unitarity might
then be logically possible.  Unitarity implies ``conservation of
microstates'', and this could lead to entropy increase much as it does
for thermodynamic systems in flat spacetime (this being essentially the
perspective adhered to by most string theorists.)  In trying to implement
such a derivation, one would have to come up with an appropriate notion
of time, to which the unitary evolution could be referred; and it is far
from clear how to do so in most black hole spacetimes.  However, it
seems to me that a more serious problem is posed by the highly
dissipative nature of the horizon.

The horizon's dissipative character is seen first of all in the ``no
hair'' theorems, according to which deviations from equilibrium
(i.e. from the Kerr metric) 
die out rapidly on a time scale set by the diameter of black hole.
In particular, this is true of the so called quasi-normal modes that
describe the ``ring down'' of a newly formed black hole, as confirmed in
numerical simulations of that process. (Indeed, the
decay is so rapid that a bell does not make a very good metaphor; a
marshmallow might be a better comparison.)  
One also recognizes dissipation in the
horizon's effective ``viscosity'' and ``electric resistivity'' 
that enter into formulas like those in [11].

Of course, one could imagine that this macroscopic relaxation to
equilibrium was balanced by the excitation of microscopic degrees of
freedom on the horizon that are invisible in the continuum picture, in
such a manner that the overall dynamics (of horizon $+$ exterior)
remained unitary.  The trouble with this idea is that semiclassical
calculations of Hawking radiation portray the radiated quanta as
strongly correlated with modes that fall into the singularity deep
within the black hole.  Unless something could {\it remove} these
correlations, no amount of subtle, non-semiclassical correlations among
the radiated quanta (or between the latter and any horizon degrees of
freedom) could restore unitarity to the exterior region [45].  
To those who dislike it, this conclusion is known as ``the black hole
information\footnote{$^\star$}%
{\raggedbottom
 The question is sometimes asked, If information is lost in black hole
 evaporation, then where does it go?  To my mind, this way of putting
 things accords too much independent reality to the idea of information,
 which is not really a ``substance'' and does not really have a location
 in general.  Perhaps the notion of information simply fails under the
 conditions one finds deep inside the black hole (i.e. in the vicinity
 of what classically would have been the singularity), where spacetime
 itself probably no longer makes sense and is superseded by some deeper,
 discrete structure.}  %
paradox'', but my own feeling is that, 
far from being paradoxical:~\footnote{$^\dagger$}
{Some references for the fifth thesis are
 33,
 36,
 38,
 45,
 39,
 46,
 33,
 11,
 15,
 20,
 30,
 29.}

\thesis{6. This non-unitarity is to be welcomed.}

In a semiclassical treatment, the breakdown of unitarity arises from
tracing out the degrees of freedom inside the black hole; that is, it arises from a
kind of {\it coarse-graining}.  But coarse-graining is exactly what one
needs to prove entropy increase.  Indeed, all proofs of the second law
that I know of invoke coarse-graining at some stage; they are forced
to do so because evolution which is strictly unitary cannot change the
statistical entropy.  In the black hole situation, we are fortunate to
have available an objectively determined way to coarse-grain, which
therefore provides us with an objective definition of entropy.  If
anything, this is a big improvement over the relatively {\it ad hoc}
types of coarse-graining that one is usually forced to employ.

What is equally important, the one-way character of the horizon affords
a relatively direct proof of entropy increase for the coarse-grained
dynamics.  Thus the non-unitarity not only gives us a natural definition
of black hole entropy, it lets us see why the generalized second law
must hold when entropy is defined in this way.  Actually, instead of
``proof'', I should have said something like ``proof scheme'', to be
filled in once the full theory of quantum gravity is available.  This
``scheme'' relies on the assumption that the one-way character of the
horizon persists in full quantum gravity, in the sense that the
exterior region (including the horizon) evolves autonomously.  If one
adds the further assumptions that energy is conserved in the full theory
and that the momentary state of the exterior region can still be defined
by an effective density matrix (at least approximately), then one can
derive an inequality that entails the non-decreasing character of the
total entropy [37].\footnote{$^\flat$}
{Some references for the sixth thesis are
 28, 
 21, 
 7,
 37,
 5.
}

\thesis{7. Restricting to the semiclassical case we prove that the free
   energy always decreases.}

The proof scheme just sketched could only come to fruition in a theory
of full quantum gravity.  However, in the context of the semiclassical
Einstein equation, it yields the more rigorously defined statement that
the free energy, $F=E-ST$, on a hypersurface $\Sigma$
decreases monotonically as $\Sigma$ moves forward in time [45].
Here, $E$ is the energy of exterior matter fields, $S$ 
their entropy, and $T$ the temperature of a quasi-stationary black
hole.\footnote{$^\star$}
{\raggedbottom
 For simplicity, I'm assuming a non-rotating, uncharged black hole.
 The energy $E$ is defined with respect to the time-translation Killing
 vector of the black hole spacetime, normalized at infinity.}   
In other words, one is doing quantum field theory in curved
spacetime and in effect treating the black hole as a ``reservoir'' for
the external matter.  When combined with the assumption that the black
hole adjusts its mass to keep the total energy constant, this yields
immediately the ``generalized second law'' for the combined system of
black hole plus matter, in the form of the inequality:
$$
  d S_{outside}  \ge  d \langle E_{outside}\rangle / T_{BH} = - dS_{BH}
\ .
$$
This results covers almost all gedanken experiments people have
``performed'' to verify the generalized second law, e.g. those in which
one lowers or drops ``boxes'' through the horizon.\footnote{$^\dagger$}
{Some references for the seventh thesis are
 33, 36, 38, 40, 43, 26, 
 27,
 45,
 37,
 24,
 47,
 4,
 56,
 13,
 3,
 53.}

\thesis{8. The entropy would be infinite without a cutoff.}

If the above account is correct, then 
the entropy $S_{total}$ of the exterior region, 
whose monotone increase is the content of the generalized second law, 
is that derived by coarse-graining away the black hole interior.  
This entropy must include both the entropy $S_{matter}$
normally attributed to exterior matter\footnote{$^\flat$}
{The definition of $S_{matter}$ may involve a further coarse-graining
 unrelated to the presence of a horizon.} 
and the entropy $S_{BH}$ normally attributed to the black hole.  
Indeed, the latter must turn out to be (to a good approximation)
$2{\pi}A$, according to the area law.

Now one contribution to $S_{total}$ comes from 
the quantum fields inhabiting the spacetime 
(in which one can presumably include the ``gravitons''), 
and an important component of this contribution is 
the ``entanglement entropy'' belonging to correlations between field
values inside and outside the horizon.  This includes first of all the
gray-body entropy carried by the quanta of the Hawking radiation, which
are entangled with modes that propagate toward the singularity, as we
have already discussed.  But it also includes a contribution from
near-horizon modes that, even in their vacuum state, exhibit
cross-horizon correlations. (Indeed, such a contribution was also
present in the entropy $S_{outside}$ discussed in conjunction with
thesis 7, but there it dropped out of consideration because only {\it
changes} in $S$ were in question.  Notice that this near-horizon entropy
may be thought of as belonging to the ``thermal atmosphere'' of the
black hole.)

Now, in a semiclassical treatment, the entanglement entropy is easily
shown to be infinite, thanks to the divergent number of near-horizon
modes.  If, then, $S_{BH}$ is actually finite,
this can only be due to some cutoff on the modes, or to some other type
of failure of the semiclassical picture.  If one introduces a cutoff at
length scale $l$ then the entanglement entropy in the leading
approximation takes the form $S{\sim}A/l^2$, a formula which once again
points to the Planck length as the fundamental discreteness scale, and
confirms that, within the semiclassical framework, there is no escape
from the need for a cutoff.  

On the other hand, the semiclassical
treatment ignores the ``back reaction'' of field fluctuations on the
metric, and in particular on the horizon itself.  There are indications
[44] that this back reaction induces fluctuations in horizon
shape on scales well above Planckian, and this might provide a loophole
in the argument that a cutoff is necessary.  Unfortunately, this
potential loophole is difficult to assess since it brings us into the
realm of quantum gravity proper.  However, it would seem that the
horizon fluctuations, even if they suppressed the entanglement entropy
at short wavelengths, would only replace it with another infinite
contribution, this time the infinite ``geometrical'' entropy of the
shape fluctuations themselves.\footnote{$^\star$}
{In fact, it's tempting to believe that this geometrical entropy
 (suitably cut off at around $l_{Planck}$) could be viewed in an
 effective continuum description as the main source of the horizon
 entropy, in accord with the geometrical nature of the area
 law itself.}  
Once again, one would need something like an underlying spacetime
discreteness at around the Planck scale in order to avoid a divergent
entropy.\footnote{$^\dagger$}
{Some references for the eighth thesis are
 35, 5, 
 39,
 42,
 44,
 48,
 23,
 4,
 25.}

\thesis{9. To understand the generalized second law requires a spacetime
           approach, not a canonical one.} 

My ninth thesis concerns not so much what the statistical mechanics of
black holes can teach us about the micro-structure of spacetime, but
rather what it can teach us about the status in quantum gravity of
spacetime itself (of four-dimensionality as opposed to three).
Specifically, I am claiming that an approach like canonical quantum
gravity, which formulates its dynamics in terms of data on a purely
spatial 3-manifold, cannot do justice to black hole thermodynamics, and
in particular to the generalized second law.

Consider for example the proof of this law that was adumbrated in the
discussion of thesis 6.  For it to go through, we must be able to
associate an effective density matrix $\hat\rho$ to the portion of the
hypersurface $\Sigma$ 
that lies 
outside the horizon, and this in turn requires
that one be able to {\it locate} the horizon.  Can one do so if one has
access only to data on a spatial slice, but not to the spacetime that
contains it (or at least to a sufficiently great portion of that
spacetime)?  To me, the task looks hopeless.\footnote{$^\flat$}
{Even for the apparent horizon, the difficulties would be similar, as
 illustrated by controversies over the interpretation of numerical
 simulations claimed to exhibit the formation of naked singularities.}  
In contrast, a spacetime (``path integral'') formulation is, relatively
speaking, free from this difficulty (as rigorously illustrated in a
somewhat different context in [8] and [9]).

The difficulty with locating the horizon given only hypersurface data
shows up even more dramatically in connection with the analysis of
claimed giant fluctuations in the entropy induced by quantum triggers to
gravitational collapse.  Here, the very presence or absence of a massive
black hole depends on a future quantum event that, apparently, has no way
even to ``register'' on the hypersurface in question.  (The
gedankenexperiment is described more fully in [46].)\footnote{$^\star$}
{Some references for the ninth thesis are
  41,
  46,
  16,
  8,
  9,
  43,
  17,
  32,
  55,
  .}

%
%

\thesis{10. Summary}

This last thesis is not an independent claim, but merely a tying
together of the previous theses.  If they are valid, then, taken
together, they speak in favor of these conclusions: the inner basis of
spacetime has a discrete structure;  the effective degrees of freedom of
the horizon, if we can identify them, will provide a clue to the nature
of this discrete structure;  a theory of quantum gravity built on this
basis will necessarily be expressed in a language much closer to that of
``histories'' or ``path integral'' formulations of quantum mechanics
than to the purely spatial (``$3+1$'') language of canonical
quantization.\footnote{$^\dagger$}
{A further conclusion, for which however I have offered no
 evidence here, is that black holes can also teach us something about
 statistical mechanics itself, specifically that the formula
 $-\tr\hat\rho\log\hat\rho$ is singled out as the ``correct'' one for
 the Gibbs entropy [46].}

\bigskip
\noindent

This research was partly supported 
by NSF grant PHY-0404646 and
by funds from the Office of Research and Computing of Syracuse University.

\ReferencesBegin

\ref [1] A.P. Balachandran, L. Chandar and Arshad Momen (1997).
``Edge states and entanglement entropy'',
\journaldata{Int. J. Mod. Phys.}{A12}{625-642}{1997}
\eprint{hep-th/9512047}

\ref [2] Jacob D. Bekenstein (1973).
``Black Holes and Entropy'',
\journaldata{Phys. Rev. D} {7} {2333-2346} {1973}

\ref [3] J.D. Bekenstein (1981).
``A universal upper bound on the entropy to energy ratio for bounded systems''
\journaldata{Phys. Rev. D}{23}{287}{1981}

\ref [4] J.D.~Bekenstein (1994).
``Do we understand black hole entropy?'',
   in {\it The Seventh Marcel Grossmann Meeting on Recent Developments in
    Theoretical and Experimental General Relativity, Gravitation and
    Relativistic Field Theories},
    proceedings of the MG7 meeting, held Stanford, July 24--30, 1994,
    edited by R.T.~Jantzen, G.~Mac~Keis\-er and R.~Ruffini
   (World Scientific 1996)
   \eprint{gr-qc/9409015}

\ref [5] Luca Bombelli, Rabinder K. Koul, Joohan Lee and Rafael D. Sorkin (1986). \lbr
``A Quantum Source of Entropy for Black Holes''
  \journaldata{Phys. Rev. D}{34}{373-383}{1986}.
\lbr
 Re\-print\-ed in: B.-L.~Hu and L.~Parker (eds.)
 {\it Quantum Theory in Curved Spacetime} 
 (World Scientific, Singapore, 1992) 

\ref [6] Luca Bombelli, Joohan Lee, David Meyer and Rafael D.~Sorkin (1987).
``Spacetime as a Causal Set'', 
  \journaldata {Phys. Rev. Lett.}{59}{521-524}{1987}

\ref [7] Max Born (1949).
{\it Natural Philosophy of Cause and Chance},
(Oxford Univ. Press, New York, 1949; or Dover, New York, 1964)

\ref [8] Graham Brightwell, {H. Fay Dowker}, {Raquel S. Garc{\'\i}a}, {Joe Henson} 
 and {Rafael D. Sorkin} (2001).
``General Covariance and the `Problem of Time' in a Discrete Cosmology'',
 in K.G.~Bowden, Ed., 	
 {\it Correlations}, 
 Proceedings of the ANPA 23 conference,
 held August 16-21, 2001, Cambridge, England 
 (Alternative Natural Philosophy Association, London, 2002), pp 1-17.
\eprint{gr-qc/0202097}

\ref [9] Graham Brightwell, Fay Dowker, Raquel S.~Garc{\'\i}a, Joe Henson and Rafael D.~Sor\-kin (2003).
``{$\,$}`Observables' in Causal Set Cosmology'',
\journaldata {Phys. Rev.~D} {67} {084031} {2003}
\eprint{gr-qc/0210061}
%

\ref [10] S.~Carlip (1995).
``Statistical Mechanics of the (2+1)-dimensional Black Hole'',
 \journaldata{Phys. Rev. D}{51}{632-637}{1995}
 \eprint{gr-qc/9409052}.
\sepref
S. Carlip (1997).
``The statistical mechanics of the three-dimensional euclidean black hole''
\journaldata{Phys. Rev. D}{55}{878-882}{1997}
\eprint{gr-qc/9606043}

\ref [11] Thibaut Damour (1978). 
``Black-hole eddy currents'',
\journaldata{Phys. Rev. D} {18} {3598-3604} {1978}

\ref [12] Djamel Dou (1999).   
 ``Causal Sets, a Possible Interpretation for the Black Hole Entropy, and Related Topics'', 
 Ph.~D. thesis (SISSA, Trieste, 1999)
 \eprint{gr-qc/0106024}
\sepref
Djamel Dou and Rafael D.~Sorkin (2003).
``Black Hole Entropy as Causal Links''.
\journaldata {Foundations of Physics}{33}{279-296}{2003}
\eprint{gr-qc/0302009}.

\ref [13]
Valery P. Frolov and Don N. Page (1993).
``Proof of the generalized second law for quasistationary semiclassical black holes''
\journaldata{Phys. Rev. Lett.}{71}{3902-3905}{1993}
\eprint{gr-qc/9302017}

\ref [14] G.W.~Gibbons and S.W.~Hawking (1977).
``Action Integrals and Partition Functions in Quantum Gravity'',
 \journaldata{Phys.~Rev. D}{15}{2738-2751}{1977}

\ref [15]
James B. Hartle (1973).
``Tidal Friction in Slowly Rotating Black Holes'',
\journaldata{Physical Review D}{8}{1010-1024}{1973}

\ref [16] C.J.~Isham (1993).
``Canonical Quantum Gravity and the Problem of Time'',
  in L.~A.~Ibort and M.~A.~Rodriguez (eds.), 
  {\it Integrable Systems, Quantum Groups, and Quantum Field Theories}
  (Kluwer Academic Publishers, London, 1993)
   pp. 157--288
  \eprint{gr-qc/9210011}

\ref [17]
Werner Israel, ``Third law of black-hole dynamics: a formulation and proof'' (1986).
 \journaldata{Phys. Rev. Lett.}{57}{397-399}{1986}

\ref [18] Ted Jacobson (1999).
``On the nature of black hole entropy'',
 in {\it General Relativity and Relativistic Astrophysics: Eighth 
 Canadian Conference}, AIP Conference Proceedings 493,
 eds. C. Burgess and R.C. Myers (AIP Press, 1999), pp. 85-97.
\eprint{gr-qc/9908031}

\ref [19] Ted Jacobson, Donald Marolf and Carlo Rovelli (2005).
``Black hole entropy: inside or out?''
  \eprint {hep-th/0501103}		
  %


\ref [20] Kostas D. Kokkotas (1999).
``Quasi-Normal Modes of Stars and Black Holes'',
{\it Living Reviews in Relativity-1999-2}
\eprint{gr-qc/9909058}

\ref [21] Ryogo Kubo (1957).
``Statistical-mechanical theory of irreversible processes. 
I. General theory and simple applications to magnetic and conduction
processes'',
\journaldata{J. Phys. Soc. Japan}{12}{570}{1957}
 %
\sepref
Ryogo Kubo (1981).
in {\it Perspectives in Statistical Physics},
edited by H.J. Ravech{\'e}
(North Holland, Amsterdam, 1981)

\ref [22] Lev D. Landau and E.M. Lifshitz (1969).
{\it Statistical physics. Part 1},
(Pergamon Press, Oxford, 1969)

\ref [23] Christoph Holzhey, Finn Larsen and Frank Wilczek (1994).
``Geometric and Renormalized Entropy in Conformal Field Theory'', 
\lbr\journaldata{Nucl. Phys. B}{424}{1994}{443-467}
\eprint{hep-th/9403108}


\ref [24] G{\"o}ran Lindblad (1975). ``Completely Positive Maps and Entropy Inequalities'',
 \journaldata{Commun. Math. Phys.}{40}{147-151}{1975}

\ref [25] Donald Marolf (2004). ``On the quantum width of a black hole horizon'',
 in 
 {\it Particle Physics and the Universe: Proceedings of the 9th Adriatic Meeting, September 2003, Dubrovnik},
 editors J. Trampetic and J. Wess (Springer, Berlin, 2005)
\lbr\eprint{hep-th/0312059, version 3}

\ref [26] Donald Marolf and Rafael D.~Sorkin (2002). ``Perfect mirrors and the self-acceler\-ating box paradox'',
\journaldata{Phys. Rev. D}{66}{104004}{2002}
\eprint{hep-th/0201255}

\ref [27] Donald Marolf and Rafael D.~Sorkin (2004). ``On the Status of Highly Entropic Objects'',
\journaldata {Phys. Rev.~D} {69} {024014} {2004} 
\eprint{hep-th/0309218}

\ref [28] Oliver Penrose (1970). 
{\it Foundations of Statistical Mechanics}
(Pergamon Press, Oxford, 1970)

\ref [29] Roger Penrose (2004).
{\it The Road to Reality}
(Jonathan Cape, London, 2004), especially \S 30.8.


\ref [30] R.H. Price (1972).
``Nonspherical perturbations of relativistic gravitational
 collapse. I. Scalar and gravitational perturbations''
\journaldata{Phys. Rev. D}{5}{2419-2438}{1972}

\ref [31] Carlo Rovelli (1996). ``Black hole entropy from loop quantum gravity'',
\journaldata{Phys. Rev. Lett.}{77}{3288-3291}{1996}
\eprint{gr-qc/9603063}

\ref [32] Stuart L. Shapiro and Saul A. Teukolsky (1991).
``Formation of naked singularities - The violation of cosmic censorship'',
\journaldata{Phys. Rev. Lett}{66}{994-997}{1991}

\ref [33] Rafael Sorkin (1979). ``On the Meaning of the Canonical Ensemble'',
  \journaldata{Int. J. Theor. Phys.}{18}{309-321}{1979}

\ref [34] Rafael D. Sorkin, Robert M. Wald and Zhang Zhen-Jiu (1981).
``Entropy of  Self-Gravitating Radiation'', 
 \journaldata{Gen. Rel. Grav.}{13}{1127-1146}{1981}

\ref [35] Rafael D. Sorkin (1983). ``On the Entropy of the Vacuum Outside a Horizon'',
  in B. Bertotti, F. de Felice and A. Pascolini (eds.),
  {\it Tenth International Conference on General Relativity and Gravitation (held Padova, 4-9 July, 1983), Contributed Papers}, 
  vol. II, pp. 734-736
  (Roma, Consiglio Nazionale Delle Ricerche, 1983)
  %

\ref [36] Rafael D. Sorkin (1983). ``Diffusion, Differential Geometry, and Black Holes'',
  in B. Bertot\-ti, F. de Felice, A. Pascolini (eds.),
  {\it Tenth International Conference on General Relativity and Gravitation (held Padova, 4-9 July, 1983), Contributed Papers}, 
  vol. II, 737-739
  (Roma, Consiglio Nazionale Delle Ricerche, 1983)

\ref [37] Rafael D. Sorkin (1986). ``Toward an Explanation of Entropy Increase in the Presence of Quantum Black Holes'',
  \journaldata {Phys. Rev. Lett.} {56} {1885-1888} {1986}


\ref [38] Rafael D. Sorkin (1986). ``Stochastic Evolution on a Manifold of States'',
  \journaldata{Annals Phys. (New York)}{168}{119-147}{1986}

\ref [39] Rafael D. Sorkin (1987). ``A Simplified Derivation of Stimulated Emission by Black Holes'', 
 \journaldata{Classical and Quantum Gravity}{4}{L149-L155}{1987}
 %

\ref [40] Rafael D.~Sorkin (1991).
``The Gravitational-Electromagnetic Noether-Operator and 
     the Second-Order En\-er\-gy Flux'' ,
  \journaldata{Proceedings of the Royal Society London A}{435}{635-644}{1991}

\ref [41] Rafael D.~Sorkin (1993).  
``Forks in the Road, on the Way to Quantum Gravity'', talk 
   given at the conference entitled ``Directions in General Relativity'',
   held at College Park, Maryland, May, 1993,
   published in
   \journaldata{Int. J. Th. Phys.}{36}{2759--2781}{1997}   
   \eprint{gr-qc/9706002}
   %
   %

\ref [42] Rafael D.~Sorkin (1994).
``Two Topics concerning Black Holes: 
   Extremality of the Energy, Fractality of the Horizon'',
   in S.A.~Fulling (ed.), 
   {\it Proceedings of the Conference on Heat Kernel Techniques and Quantum Gravity, held Winnipeg, Canada, August, 1994}, pp. 387-407
   (Discourses in Mathematics and its Applications, \#4) 
   (University of Texas Press, 1995)
   \eprint{gr-qc/9508002}

\ref [43] Rafael D.~Sorkin and Madhavan Varadarajan (1996). ``Energy Extremality in the Presence of a Black Hole'',
  \journaldata {Class. Quant. Grav.} {13} {1949-1970} {1996}
  \eprint{gr-qc/9510031}
 %

\ref [44] Rafael D.~Sorkin (1996).
``How Wrinkled is the Surface of a Black Hole?''.
  in David Wiltshire (ed.), 
  {\it Proceedings of the First Australasian Conference on General Relativity and Gra\-vitation}, 
  held February 1996, Adelaide, Australia, pp. 163-174
  (University of Adelaide, 1996)
  \eprint{gr-qc/9701056}



\ref [45] Rafael D.~Sorkin (1998). ``The Statistical Mechanics of Black Hole Thermodynamics'',
  in R.M. Wald (ed.) {\it Black Holes and Relativistic Stars}, 
  (U. of Chicago Press, 1998), pp. 177-194
  \eprint{gr-qc/9705006}


%

\ref [46] Rafael D.~Sorkin and Daniel Sudarsky (1999).
``Large Fluctuations in the Horizon Area and What They Can Tell Us About
   Entropy and Quantum Gravity''
  \journaldata {Class. Quant. Grav.}{16}{3835-3857}{1999}
  \eprint {gr-qc/9902051}		


\ref [47] Herbert Spohn (1978).
``Entropy production for quantum dynamical semigroups''
\journaldata{Journal of Mathematical Physics}{19}{1227-1230}{1978}
\sepref
H. Spohn and J.L. Lebowitz (1978).
``Irreversible thermodynamics for quantum systems weakly coupled to thermal reservoirs''
\journaldata{Adv. Chem. Phys.}{38}{109-142}{1978}

\ref [48] Mark Srednicki (1993).
``Entropy and Area'',
 \journaldata{Phys.~Rev.~Lett.}{71}{666-669}{1993}
 \eprint{hep-th/9303048}

\ref [49] Leonard Susskind and John Uglum (1994).	      
``Black hole entropy in canonical quantum gravity and superstring theory'',
  \journaldata{Phys. Rev. D}{50}{2700-2711}{1994}
  \eprint{hep-th/9401070}
  %

\ref [50] L. Susskind (1995).
``The world as a hologram'',
\journaldata{J. Math. Phys.}{36}{6377}{1995}
\lbr\eprint{hep-th/9409089}

\ref [51] G.~'t~Hooft (1985). ``On the quantum structure of a black hole'', 
   \journaldata{ Nuclear Phys. B}{256}{727-745}{1985}
  %

\ref [52] Kip S. Thorne, Richard H. Price and Douglas A. Macdonald (eds.) (1986).
  {\it Black Holes: The Membrane Paradigm}
  (Yale University Press, New Haven, 1986)

\ref [53] W.G. Unruh and R.M. Wald (1982).
``Acceleration radiation and generalized second law of thermodynamics'',
\journaldata{Phys. Rev. D}{25}{942}{1982} 

\ref [54] Robert M. Wald (1998).
``Black Holes and Thermodynamics''
  in R.M. Wald (ed.) {\it Black Holes and Relativistic Stars}, 
  (U. of Chicago Press, 1998), pp. 155-176

\ref [55] Robert M. Wald and Vivek Iyer (1991).
``Trapped surfaces in the Schwarzschild geometry and cosmic censorship
 \journaldata{Phys. Rev. D}{44}{R3719-R3722}{1991}

\ref [56] W.H.~Zurek and K.S.~Thorne (1985).
``Statistical Mechanical Origin of the Entropy of 
  a Rotating, Charged Black Hole'',
 \journaldata{Phys. Rev. Lett.}{54}{2171-2175}{1985}

\end     



  (Outline*
   "
   "\\message"   1
   "\\Abstract"   1
   "\\thesis"    1
   "\\Referen"   1
   "\\ref"       2
   "\\end